\documentclass[12pt]{article}

\usepackage{amsmath,amssymb}

\global\arraycolsep=1pt
\numberwithin{equation}{section}
\newcommand{\bel}[1]{\begin{equation}\label{#1}}                     
\newcommand{\bal}[1]{\begin{eqnarray}\label{#1}}                     
\newcommand{\be}{\begin{equation}}
\newcommand{\ee}{\end{equation}}

\newcommand{\im}{\mathrm{i}}

\newcommand{\de}{\mathrm{d}}

\newcommand{\qq}{\qquad}
\newcommand{\mat}[1]{\begin{pmatrix} #1 \end{pmatrix}}
\renewcommand{\thefootnote}{\fnsymbol{footnote}}
\newcommand{\ul}[1]{\underline{#1}}
\newcommand{\ol}[1]{\overline{#1}}

\begin{document}

%
%
\begin{titlepage}

\begin{flushright}
\normalsize
~~~~
October, 2006 \\
OCU-PHYS 257 \\
hep-th/0610325 \\
\end{flushright}

\begin{center}
{\large\bf The $AdS_5 \times S^5$ superstrings \\
  in the generalized light-cone gauge }
\end{center}

\vfill

\begin{center}
{%
H. Itoyama$^a$$^b$\footnote{e-mail: itoyama@sci.osaka-cu.ac.jp}
\quad and \quad
T. Oota$^b$\footnote{e-mail: toota@sci.osaka-cu.ac.jp}
}
\end{center}

\vfill

\begin{center}
$^a$ \it Department of Mathematics and Physics,
Graduate School of Science\\
Osaka City University\\
\medskip

$^b$ \it Osaka City University Advanced Mathematical Institute
(OCAMI)

\bigskip

3-3-138, Sugimoto, Sumiyoshi-ku, Osaka, 558-8585, Japan \\

\end{center}

\vfill


\begin{abstract}
The $\kappa$-symmetry-fixed Green-Schwarz action
in the $AdS_5 \times S^5$ background is treated canonically 
in a version of the light-cone gauge.
After reviewing the generalized light-cone gauge
for a bosonic sigma model, we present the
Hamiltonian dynamics of the Green-Schwarz action
by using the transverse degrees of freedom.
The remaining fermionic constraints are all second class,
which we treat by the Dirac bracket.
Upon quantization, 
all of the transverse coordinates are inevitably non-commutative. 
\end{abstract}

\vfill

\end{titlepage}

\renewcommand{\thefootnote}{\arabic{footnote}}
\setcounter{footnote}{0}


\section{Introduction}

Since the proposal of AdS/CFT correspondence,
it has become an important issue 
to quantize the type IIB Green-Schwarz superstrings \cite{GS,GS2}
in the $AdS_5 \times S^5$ background \cite{MT}.
One of the difficulties in quantizing the Green-Schwarz superstrings
stems from the existence of the local $\kappa$ symmetry,
which halves the fermionic degrees of freedom.
In the canonical Hamiltonian formalism,
the local $\kappa$ symmetry yields 
fermionic constraints. 
The half of these are first-class and
the remaining half are second-class constraints.
Covariant separation of the first and the second class 
constraints is a difficult task \cite{GS,GS2,HK}.

In the flat Minkowski target space, 
there was an attempt to quantize the action covariantly by
introducing an infinite number of ghosts (see for example \cite{bri}).
Other direction for covariant quantization is to add
extra degrees of freedom 
in order to replace the second-class constraints
with the first-class ones \cite{sie,ber,ber2,OT,GPvN,AK,GG}.

A less ambitious way to quantize the Green-Schwarz action
is to abandon the covariance and 
to go to  a non-covariant gauge.
In flat target space, 
the Green-Schwarz action in the light-cone
gauge becomes extremely simple \cite{GS,GS2}.
Light-cone quantization of quantum field theory 
was first recognised and developed  in connection with 
the current algebra in the infinite momentum
frame \cite{SUSS,BH,KS}.
Light-cone quantization of (super)-strings played important
roles in the development of string theory in seventies 
\cite{MAND,GGRT,GS73,KK} and that of superstring theory in
eighties \cite{GS,GS2}.
Various gauges for the $AdS_5 \times S^5$ superstrings
have been proposed \cite{pes,kal,KR98,KT98,MT00,MTT00}.

Recently, the Hofman-Maldacena limit \cite{HM} 
has attracted much attention \cite{dor,MW,RTT,GK,MPPSZ,CDO,
BC,AFZ,MTT06,bei06,CGK,SV,BR1,KRT06,BR2,hua,vaz,BD,
GH,roi,CS,OS,AFPZ,hir,rya,SZZ}. 
It is a limit which takes the energy $E$ and
one of the angular momenta $J$ infinite while keeping $E-J$ finite.
$E$ and $J$ are eigenvalues of Cartan generators
of $SO(2,4)$ and $SO(6)$ group respectively.
In the Hofman-Maldacena limit, 
both the string and the dual gauge theory
describe excitations called giant magnons and
their generalizations.
Good agreement in some physical quantities
is found.

One way to take the Hofman-Maldacena limit is to
employ a version of light-cone gauge
in which $E-J$ appears as the light-cone energy.
A sizable amount of literature
has been accumulated which are devoted to the discussion
of this gauge \cite{KRT04,KT04,AF04,AF05,MTT00,FPZ,AFPZ}.
This gauge is sometimes referred to as the uniform light-cone gauge
and is a generalization of that of
\cite{GGRT} in the flat Minkowski space to the AdS background.
The light-cone direction $X^{\pm}$ is chosen
such that $X^{\pm} = (1/\sqrt{2})(t \pm \varphi)$,
where $t$ is the global time direction of $AdS_5$
and $\varphi$ is a certain angle of $S^5$. 
The vectors $\partial/\partial X^{\pm}$ are Killing vectors
of the target space geometry.
The transverse direction manifestly 
keeps the covariance under a $SO(4) \times SO(4)$ 
subgroup of the local Lorentz group 
$SO(1,4) \times SO(5)$ \cite{FPZ}.
The treatment of the fermionic second class constraints 
remains to be investigated however.
In order to treat these remaining constraints, 
it is necessary to introduce the Dirac bracket.

In this paper, 
we study the $AdS_5 \times S^5$ superstring
in the generalized light-cone gauge 
as a constrained Hamiltonian system. 
In section $2$, we review the generalized light-cone gauge for
the bosonic sigma models,
emphasizing the central object, the light-cone Hamiltonian.
In section $3$, we study the case 
of the Green-Schwarz superstring.
The fermionic second-class constraints 
lead to the highly non-trivial
Dirac bracket among the transverse degrees of freedom.
If  $\im \hbar$ times the Dirac bracket is replaced
with the graded commutator, all of the transverse
coordinates are inevitablly non-commutative.
Since the Dirac brackets are not $c$-number, several subtleties
such as operator ordering remain to be investigated. 
In the Appendix, we give some details on the induced vielbeins.


\section{Bosonic sigma models in the generalized light-cone gauge}


\subsection{Bosonic sigma model}

In order to explain the generalized light-cone gauge,
let us consider the following bosonic sigma model:
\be
S = \frac{1}{2\pi} \int \de^2 \xi \, \mathcal{L},
\ee
where the Lagrangian density is given by
\bel{BSL}
\mathcal{L} = 
- \frac{1}{2} \sqrt{\lambda}
h^{i j} G_{\ul{m}\, \ul{n}}(X) \partial_{i} X^{\ul{m}}
\partial_{j} X^{\ul{n}}.
\ee
We assume that the target space is $D$-dimensional:
$X^{\ul{m}} = X^{\ul{m}}(\xi)$, 
$(\ul{m}=0,1,\dotsc, D-1)$,
and $G_{\ul{m} \, \ul{n}}(X)$ is the metric of the target space.
Here 
\be
(\xi^0, \xi^1 ) = (\tau, \sigma), \qq
h^{ij} = \sqrt{-g} g^{ij}, \qq
i,j=0,1,
\ee
and $\lambda$ is the coupling constant.
$h^{ij}$ is the Weyl-invariant combination of the world-sheet
metric $g_{ij}$.
Since $\det h^{ij}= -1$, we choose $h^{00}$ and $h^{01}$
as the independent Lagrange multipliers.
We also use the following notation for the Lagrange
multipliers: $e^0 = 1/(2h^{00})$, $e^1=(h^{01}/h^{00})$.

Equations of motion for this model are given by
\bel{EOM1}
\partial_{i} ( h^{i j} G_{\ul{m}\, \ul{n}} \partial_{j} X^{\ul{n}} )
= \frac{1}{2} h^{i j} 
( \partial_{\ul{m}} G_{\ul{k}\, \ul{l}} ) \partial_{i} X^{\ul{k}} 
\partial_{j} X^{\ul{l}}.
\ee
Here 
$\partial_{\ul{m}} = \partial/\partial X^{\ul{m}}$.

Let us introduce the conjugate momenta by
$P_{\ul{m}}:= \partial \mathcal{L}/\partial \dot{X}^{\ul{m}}$.
\bel{PM}
P_{\ul{m}} = - \sqrt{\lambda} G_{\ul{m}\, \ul{n}} 
h^{0 i} \partial_{i} X^{\ul{n}}.
\ee
Let $G^{\ul{m}\, \ul{n}}$ be the inverse of the 
metric $G_{\ul{m}\, \ul{n}}$.
The equations of motion \eqref{EOM1} and the definition
of the conjugate momenta \eqref{PM} can be converted into
the equations of motion in the first order form:
\bel{EOM2}
\begin{split}
\dot{X}^{\ul{m}} 
&= - \frac{1}{\sqrt{\lambda} h^{0 0}} 
G^{\ul{m}\, \ul{n}} P_{\ul{n}}
- \left( \frac{h^{0 1}}{h^{0 0}} \right) 
\partial_{1} X^{\ul{m}}, \cr
\dot{P}_{\ul{m}} &= \partial_{1}
\left[ - \left( \frac{h^{0 1}}{h^{0 0}} \right)
P_{\ul{m}} - \frac{\sqrt{\lambda}}{h^{0 0}} G_{\ul{m}\, \ul{n}} 
\partial_{1} X^{\ul{n}} \right] \cr
& + \frac{\sqrt{\lambda}}{2 h^{0 0}}
\left[ \frac{1}{\lambda} 
( \partial_{\ul{m}} G^{\ul{k}\, \ul{l}} ) P_{\ul{k}} P_{\ul{l}}
+ ( \partial_{\ul{m}} G_{\ul{k}\, \ul{l}}) 
\partial_{1} X^{\ul{k}} \partial_{1} X^{\ul{l}} \right].
\end{split}
\ee

The Hamiltonian density is given by 
\be
\mathcal{H}= P_{\ul{m}} \dot{X}^{\ul{m}} 
- \mathcal{L} 
= - e^0 \Phi_0 - e^1 \Phi_1,
\ee
where
\be
\Phi_0:= \frac{1}{\sqrt{\lambda}} G^{\ul{m}\, \ul{n}} 
P_{\ul{m}} P_{\ul{n}} 
+ \sqrt{\lambda} G_{\ul{m}\, \ul{n}} \partial_{1} X^{\ul{m}}
\partial_{1} X^{\ul{n}},
\qq
\Phi_1:= P_{\ul{m}} \partial_{1} X^{\ul{m}}.
\ee
The Virasoro constraints are given by
$\Phi_0 \approx 0$, $\Phi_1 \approx 0$.
The Hamiltonian density vanishes weakly:
$\mathcal{H} \approx 0$.
Using \eqref{EOM2}, we can check that
the Virasoro constraints are consistent with the time evolution
\be
\begin{split}
\dot{\Phi}_0 &= - 2 (\partial_{1} e^1) \Phi_0
- 8  ( \partial_{1} e^0) \Phi_1
- \partial_{1} \Phi_0 
- 2 \partial_{1} \Phi_1, \cr
\dot{\Phi}_1 &= 
- 2 ( \partial_{1} e^0) \Phi_0 - 2 (\partial_{1} e^1) \Phi_1
- e^0  \partial_{1} \Phi_0
- e^1 \partial_{1} \Phi_1.
\end{split}
\ee
There is no secondary constraint.


\subsection{Generalized light-cone gauge}

Let us decompose the target space index $\ul{m}$ into
$\ul{m} = (\mathfrak{a}, m)$, 
$\mathfrak{a}=\pm$, 
$m=1,2,\dotsc, D-2$.
We assume that the target space metric takes the form
\be
G_{\ul{m}\, \ul{n}} \de X^{\ul{m}} \de X^{\ul{n}} 
= G_{\mathfrak{a}\mathfrak{b}} \de X^{\mathfrak{a}} 
\de X^{\mathfrak{b}} 
+ G_{mn} \de X^m \de X^n,
\ee
and $\partial/\partial X^{\pm}$ are Killing vectors.

We first recall the procedure of the light-cone gauge fixing
in the flat target space. In this case,
the world-sheet diffeomorphism is fixed
by setting the world-sheet metric conformally flat.
The residual symmetry is used to set
$X^+ = \kappa \tau$.

But in a certain curved target space
such as $AdS$ space-time, there is an obstacle in
making the world-sheet metric be conformally flat
and obey the light-cone gauge condition \cite{MTT00}. 
Instead, in the generalized light-cone approach,
the world-sheet 
diffeomorphism is fixed by imposing 
the following two conditions\footnote{Here for simplicity 
we consider the sector with vanishing winding number.}:
\be
X^+ = \kappa \tau, \qq \dot{P}_- = 0.
\ee
Equation of motion \eqref{EOM2} for $X^+$
\be
\dot{X}^+ = \kappa = - \frac{1}{\sqrt{\lambda} h^{0 0}}
( G^{++} P_+ + G^{+-} P_- )
\ee
determines the Lagrange multiplier $h^{00}$ as
\be
h^{0 0} = - \frac{1}{\sqrt{\lambda} \kappa}
( G^{++} P_+ + G^{+-} P_- ).
\ee
Equation of motion for $P_-$
\be
0 = \dot{P}_- = \partial_{1}
\left[ - \left(
\frac{h^{0 1}}{h^{0 0}} \right)
P_- - \frac{\sqrt{\lambda}}{h^{0 0}} 
G_{--} \partial_{1} X^- \right]
\ee
fixes $h^{0 1}$ up to an arbitrary function of $\tau$:
\be
h^{0 1} = - \frac{\sqrt{\lambda}}{P_-} G_{--}
\partial_{1} X^- - \frac{f(\tau)}{P_-} h^{0 0}.
\ee
The function $f(\tau)$ arises from the residual symmetry.
The residual symmetry is fixed by setting $f(\tau) = 0$.
Solving the Virasoro constraint $\Phi_1=0$ gives the relation
$\partial_{1} X^- = - (1/P_-) P_m \partial_{1} X^m$.

Therefore, the worldsheet metric is fixed as
\be
h^{0 0} = - \frac{1}{\sqrt{\lambda} \kappa}
( G^{++} P_+ + G^{+-} P_- ), \qq
h^{0 1} = \frac{\sqrt{\lambda}}{P_-^2} G_{--}
P_m \partial_{1} X^m.
\ee
The Virasoro constraint $\Phi_0=0$ gives a quadratic equation for $P_+$:
\bel{qua1}
G^{++} P_+^2 + 2 P_- G^{+-} P_+ + P_-^2 G^{--}
+ G^{mn} P_m P_n + \frac{\lambda}{P_-^2} G_{--}
( P_m \partial_{1} X^m )^2
+ \lambda G_{mn} \partial_{1} X^m \partial_{1} X^n = 0.
\ee

The equations of motion for the dynamical variables in the
reduced phase space  are given by
\be
\begin{split}
\dot{X}^m &= - \frac{\sqrt{\lambda}}{h^{0 0}}
\left[ \frac{1}{\lambda} G^{mn} P_n + \frac{G_{--}}{P_-^2}
( P_n \partial_{1} X^n ) \partial_{1} X^m \right], \cr
\dot{P}_m &= \partial_{1}\left[
- \left( \frac{h^{0 1}}{h^{0 0}} \right) P_m
- \frac{\sqrt{\lambda}}
{h^{0 0}} G_{mn} \partial_{1} X^n \right] \cr
& + \frac{1}{2 \sqrt{\lambda}h^{0 0}}
\left[ \frac{}{} (\partial_m G^{++}) P_+^2 + 2 (\partial_m G^{+-}) P_+ P_-
+ ( \partial_m G^{--}) P_-^2 + ( \partial_m G^{kl}) P_k P_l \right. \cr
& \qq \qq \qq
\left. + 
\frac{\lambda}{P_-^2} ( \partial_m G_{--}) (P_n \partial_{1} X^n)^2
+ \lambda ( \partial_m G_{kl}) 
\partial_{1} X^k \partial_{1} X^l \right].
\end{split}
\ee
Using the Poisson bracket
$\{ X^m(\tau, \sigma), P_n(\tau, \sigma') \}_{\mathrm{P.B.}}
= 2\pi \delta^m_n \delta(\sigma - \sigma')$,
the equations of motion can be rewritten as
\be
\dot{X}^m(\tau, \sigma) = 
\{ X^m(\tau, \sigma), H_{\mathrm{LC}} \}_{\mathrm{P.B.}}, \qq
\dot{P}_m(\tau, \sigma) = 
\{ P_m(\tau, \sigma), H_{\mathrm{LC}} \}_{\mathrm{P.B.}}.
\ee
The light cone Hamiltonian is found to be\footnote{
In addition we must examine the open string and the closed string
boundary conditions and the level-matching condition.
We will not dwell upon these in this paper.}
\be
H_{\mathrm{LC}}:= - \frac{\kappa}{2\pi} \int_{-\pi}^{\pi} \de \sigma \,  P_+.
\ee
Here $P_+$ is a solution to the quadratic equation \eqref{qua1}.

In the Hamilton formalism, the light-cone Hamiltonian $H_{\mathrm{LC}}$
can be understood by using a canonical transformation
$(X^{\ul{m}}, P_{\ul{m}}) \rightarrow (\tilde{X}^{\ul{m}}, \tilde{P}_{\ul{m}})$
\be
\tilde{X}^+ = X^+ - \kappa \tau, \qq
\tilde{X}^- = X^-, \qq \tilde{X}^m = X^m, \qq
\tilde{P}_{\ul{m}} = P_{\ul{m}},
\ee
whose generating functional is given by
\be
W(X, \tilde{P}, \tau ) = 
\int \frac{\de \sigma}{2\pi} \left[
 ( X^+(\tau,\sigma) - \kappa \tau ) \tilde{P}_+(\tau, \sigma)
+ X^-(\tau, \sigma) \tilde{P}_-(\tau, \sigma)
+ X^m(\tau, \sigma) \tilde{P}_{m}(\tau, \sigma) \right].
\ee
The transformed Hamiltonian is given by
\be
\tilde{H} = H + \frac{\partial W}{\partial \tau}
= H_{LC}.
\ee

 
\subsection{$AdS_5 \times S^5$ case}

The bosonic part of the Green-Schwarz model
for the $AdS_5 \times S^5$ background is
a special case of the sigma model \eqref{BSL}.
The coordinates for $D=10$-dimensional target space is chosen as
\be
X^{\ul{m}} = ( X^+, X^-, X^a, X^{4+s} ), \qq
X^a = z^a, \ \ a=1,2,3,4, \qq
X^{4+s} = y^s, \ \ s=1,2,3,4,
\ee
and the $AdS_5 \times S^5$ metric is given by
\bel{metricAdS5andS5}
G_{\ul{m} \, \ul{n}}(X)
\de X^{\ul{m}} \de X^{\ul{n}}
= G_{\mathfrak{a} \mathfrak{b}}
\de X^{\mathfrak{a}} \de X^{\mathfrak{b}}
+ G_z \sum_{a=1}^4 ( \de z^a)^2 + G_y \sum_{s=1}^4 ( \de y^s)^2,
\ee
where
\be
G_{++} = G_{--} = -\frac{1}{2} 
\left( \frac{1 + (z^2/4)}{1-(z^2/4)} \right)^2
+ \frac{1}{2} \left( \frac{1-(y^2/4)}{1+(y^2/4)} \right)^2,
\ee
\be
G_{+-} = G_{-+} = -\frac{1}{2} 
\left( \frac{1 + (z^2/4)}{1-(z^2/4)} \right)^2
- \frac{1}{2} \left( \frac{1-(y^2/4)}{1+(y^2/4)} \right)^2,
\ee
\be
G_z = \frac{1}{(1-(z^2/4))^2}, \qq
G_y= \frac{1}{(1+(y^2/4))^2}.
\ee
Here
\be
z^2 = \sum_{a=1}^4 (z^a)^2, \qq
y^2 = \sum_{s=1}^4 (y^s)^2.
\ee
The coupling constant $\lambda$
is related to the radius $R$ of the $AdS_5$
and $S^5$ as follows: $\sqrt{\lambda} = R^2/\alpha'$.

In the generalized light-cone gauge, $P_+$ is determined
by the following equation
\be
G^{++} P_+^2 + 2 B P_+ + C = 0,
\ee
where $B = G^{+-} P_-$,
\be
\begin{split}
C &= G^{--} P_-^2 + \frac{1}{G_z} \sum_{a=1}^4 P_a^2
+ \frac{1}{G_y} \sum_{s=1}^4 P_{4+s}^2 \cr
&+ \frac{\lambda}{P_-^2} G_{--} ( P_a \partial_1 z^a 
+ P_{4+s} \partial_1 y^s )^2
+ \lambda G_z \sum_{a=1}^4 (\partial_1 z^a)^2
+ \lambda G_y \sum_{s=1}^4 (\partial_1 y^s)^2.
\end{split}
\ee

For $AdS_5 \times S^5$, we can take the flat Minkowski limit
$R\rightarrow \infty$.
In this case,
\be
G^{++} = 0 + O(R^{-2}), \qq
G^{+-} = -1 + O(R^{-2}).
\ee
Therefore, in order to have a finite Minkowski limit,
the sign for $P_+$ must be chosen as
\be
P_+ = \frac{1}{G^{++}}
( - B + \epsilon_B \sqrt{B^2 - G^{++} C}),
\ee
where
$\epsilon_B$ is $1$ for $B>0$ and $-1$ for $B<0$.


\section{The $AdS_5 \times S^5$ Green-Schwarz superstring
in the generalized light-cone gauge}


\subsection{The Green-Schwarz action in the $AdS_5 \times S^5$ background}

The Green-Schwarz superstring in the flat target space
was proposed in \cite{GS,GS2}. Generalization 
to the action for the curved supergravity backgroud 
was done in \cite{GHMNT}.

More explicit Green-Schwarz action 
in the $AdS_5 \times S^5$ background was constructed
in \cite{MT} based on the coset superspace
$PSU(2,2|4)/( SO(1,4) \times SO(5))$. (See also \cite{KRR,MT2}).
Originally, the Wess-Zumino term is written in the
three-dimensional form. 
The manifestly two-dimensional
form of the Wess-Zumino term was presented in \cite{BBHZZ,ber,RS}.

The Green-Schwarz action for the $AdS_5 \times S^5$
is given by
\be
S_{\mathrm{GS}} = 
\frac{1}{2\pi} \int \de^2 \xi \, \mathcal{L}_{\mathrm{GS}},
\ee
\bel{LGSAdS}
\mathcal{L}_{\mathrm{GS}} = 
- \frac{1}{2} \sqrt{\lambda} h^{ij} 
\, \eta_{\ul{a} \, \ul{b}}
E^{\ul{a}}_i E^{\ul{b}}_j
+ \sqrt{\lambda} \epsilon^{ij}
\bigl( E^{\ul{\alpha}}_i \varrho_{\ul{\alpha} \ul{\beta}} E^{\ul{\beta}}_j
- \ol{E}^{\ul{\bar{\alpha}}}_i 
\varrho_{\ul{\bar{\alpha}} \ul{\bar{\beta}}} 
\ol{E}^{\ul{\bar{\beta}}}_j \bigr). 
\ee
Here $E^A_i$ 
is the induced vielbein for the type IIB superspace:
\be
E^A_i = E^A{}_M \partial_i Z^M
= E^A{}_{\ul{m}} \partial_i X^{\ul{m}}
+ E^A{}_{\ul{\mu}} \partial_i \theta^{\ul{\mu}} 
+ \ol{E}^A{}_{\ul{\bar{\mu}}} \partial_i 
\bar{\theta}^{\ul{\bar{\mu}}}.
\ee
The local Lorentz index $A=(\ul{a}, \ul{\alpha}, \ul{\bar{\alpha}})$
take values in the following way:
$\ul{a}=(\mathfrak{a}, a, 4+s)$,
$\mathfrak{a}=\pm$, 
$a=1,2,3,4$, 
$s=1,2,3,4$,
$\ul{\alpha}=1,2,\dotsc, 16$ and 
$\ul{\bar{\alpha}}
= \bar{1}, \bar{2},\dotsc, \bar{16}$.
We use the $16$-component notation for Weyl spinors.
The constant matrix $\varrho$ in the Wess-Zumino term
is given by
\be
C \Gamma^{01234} = \mat{ \varrho_{\ul{\alpha} \ul{\beta}}
& \ & 0 \cr
0 & & \varrho^{\ul{\alpha} \ul{\beta}} },
\qq \Gamma^{\ul{a}} = \mat{ 0 & \ & 
( \gamma^{\ul{a}} )^{\ul{\alpha} \ul{\beta}} \cr
( \gamma^{\ul{a}} )_{\ul{\alpha} \ul{\beta}} & & 0 }.
\ee
It is related to the existence of the self-dual
Ramond-Ramond $5$-form flux. 

In the large radius limit, \eqref{LGSAdS}
goes to the Lagrangian in the flat Minkowski space 
up to (divergent) surface terms\footnote{The surface terms
purely come from the Wess-Zumino term.}.

Let us decompose each of the two $16$-component Weyl spinors 
into two $8$-component $SO(4) \times SO(4)$ spinors:
\be
\theta^{\ul{\alpha}} = \mat{ \theta^{+\alpha} \cr
\theta^{-{\dot{\alpha}}} }, \qq
\bar{\theta}^{\ul{\bar{\alpha}}}
= \mat{ \bar{\theta}^{+\bar{\alpha}} \cr
\bar{\theta}^{- \dot{\bar{\alpha}}} },
\ee
where 
$\alpha=1,2,\dotsc, 8$, 
$\dot{\alpha} = \dot{1}, \dot{2}, \dotsc, \dot{8}$,
$\bar{\alpha}=\bar{1},\bar{2},\dotsc, \bar{8}$ and
$\dot{\bar{\alpha}}=\dot{\bar{1}}, \dot{\bar{2}}, \dotsc, 
\dot{\bar{8}}$.

We first fix the $\kappa$-symmetry by setting
$\theta^{-\dot{\alpha}} = 
\bar{\theta}^{-\dot{\bar{\alpha}}} = 0$.
In the $32$-component notation, these conditions are 
equivalent to the condition $\Gamma^+ \Theta = 0$.
In the large radius limit, it directly goes
to the $\kappa$-symmetry fixing condition for the flat
Minkowski target space.

To simplify expressions, we combine the remaining fermionic
coordinates into $\Psi^{\hat{\alpha}}$:
\be
(\Psi^{\hat{\alpha}}) 
= \mat{ \theta^{+\alpha} \cr \bar{\theta}^{+ \bar{\alpha}} },
\qq \hat{\alpha}=\hat{1}, \hat{2}, \dotsc, \hat{16}.
\ee
The coordinates for the reduced type IIB superspace is given by
$Z^M = ( X^{\ul{m}}, \Psi^{\hat{\alpha}} )
= ( X^+, X^- , X^m, \theta^{+\alpha}, \bar{\theta}^{+\bar{\alpha}})$.
We further decompose $X^m = (X^a, X^{4+s}) = (z^a, y^s)$
and choose a representative of the coset superspace as follows
\bel{CRG}
G(Z) = \exp\left( X^+ \widehat{P}_+ + X^- \widehat{P}_-  \right)
\exp\left( \theta^{+\alpha} \widehat{Q}_{\alpha}^+ 
+ \bar{\theta}^{+\bar{\alpha}} \widehat{\ol{Q}}_{\bar{\alpha}}
\! \! \! {}^+ \right)
g_{z} g_y.
\ee
Here $\widehat{P}_{\pm}$, $\widehat{Q}_{\alpha}^+$
and $\widehat{\ol{Q}}{}_{\bar{\alpha}}^+$
belong to $psu(2,2|4)$ generators. The vielbeins $E^A{}_M$
can be read from the
Cartan one-form $G^{-1} \de G$.
See Appendix for details.

The $\kappa$-symmetry fixed action for $AdS_5 \times S^5$
can be written as
\bel{KFGS}
\mathcal{L}_{\mathrm{GS}}
= - \frac{1}{2} \sqrt{\lambda} h^{ij} G_{\ul{m}\, \ul{n}}(X)
\mathcal{D}_i X^{\ul{m}} \mathcal{D}_j X^{\ul{n}}
+ \frac{1}{2} \sqrt{\lambda} \epsilon^{ij}
B_{\hat{\alpha} \hat{\beta}} \mathcal{D}_i \Psi^{\hat{\alpha}}
\mathcal{D}_j \Psi^{\hat{\beta}}.
\ee
The target space metric $G_{\ul{m}\, \ul{n}}$
is the same as the bosonic one \eqref{metricAdS5andS5},
$B_{\hat{\alpha} \hat{\beta}} = B_{\hat{\alpha} \hat{\beta}}(Z)$,
and $\Lambda$'s are introduced through
\bel{covD}
\begin{split}
\mathcal{D}_i X^+ &= \partial_i X^+, \cr
\mathcal{D}_i X^- &= \partial_i X^- 
+ \Lambda^-{}_{\hat{\alpha}} \mathcal{D}_i \Psi^{\hat{\alpha}}, \cr
\mathcal{D}_i X^m &= \partial_i X^m 
+ ( \Lambda^m{}_{n \hat{\alpha}} 
\mathcal{D}_i \Psi^{\hat{\alpha}} ) X^n , \cr
\mathcal{D}_i \Psi^{\hat{\alpha}}
&= \partial_i \Psi^{\hat{\alpha}} 
+ ( \Lambda^{\hat{\alpha}}{}_{\hat{\beta}}
\partial_i X^+ ) \Psi^{\hat{\beta}}.
\end{split}
\ee
Here $\Lambda^-{}_{\hat{\alpha}}$ and
$\Lambda^m{}_{n \hat{\alpha}}$ depend only on fermionic variables
$\Psi^{\hat{\gamma}}$ and 
$\Lambda^{\hat{\alpha}}{}_{\hat{\beta}}$ is a constant.
See Appendix for details.

The conjugate momenta are given by
\bel{conjM}
\begin{split}
P_+ &= - \sqrt{\lambda} h^{0i} 
G_{+,\mathfrak{a}} \mathcal{D}_i X^{\mathfrak{a}}
+ \mathcal{P}_{\hat{\alpha}} 
\Lambda^{\hat{\alpha}}{}_{\hat{\beta}} \Psi^{\hat{\beta}}, \cr
P_- &= - \sqrt{\lambda} h^{0i} G_{-,\mathfrak{a}} 
\mathcal{D}_i X^{\mathfrak{a}}, \cr
P_m &= - \sqrt{\lambda} h^{0i} G_{mn} \mathcal{D}_i X^n, \cr
P_{\hat{\alpha}}
&= - \sqrt{\lambda} B_{\hat{\alpha} \hat{\beta}} 
\mathcal{D}_1 \Psi^{\hat{\beta}}
+ P_- \Lambda^-{}_{\hat{\alpha}} 
+ P_m \Lambda^{m}{}_{n \hat{\alpha}} X^n.
\end{split}
\ee
We have fermionic primary constraints:
\bel{PCF}
\Phi_{\hat{\alpha}}
= P_{\hat{\alpha}} + \sqrt{\lambda} B_{\hat{\alpha} \hat{\beta}}
\mathcal{D}_1 \Psi^{\hat{\beta}}
- P_- \Lambda^-{}_{\hat{\alpha}} - P_m \Lambda^m{}_{n \hat{\alpha}}
X^n \approx 0.
\ee
The Hamiltonian density is given by
\be
\mathcal{H} = P_{\ul{m}} \dot{X}^{\ul{m}} + P_{\hat{\alpha}}
\dot{\Psi}^{\hat{\alpha}}
- \mathcal{L} 
= - e^0 \, \Phi_0 
- e^1 \Phi_1,
\ee
where 
\be
\Phi_0 = \frac{1}{\sqrt{\lambda}} G^{\mathfrak{a} \mathfrak{b}}
\Pi_{\mathfrak{a}} \Pi_{\mathfrak{b}}
+ \sqrt{\lambda} G_{\mathfrak{a} \mathfrak{b}}
\mathcal{D}_1 X^{\mathfrak{a}}
\mathcal{D}_1 X^{\mathfrak{b}} \
 + \frac{1}{\sqrt{\lambda}} G^{mn} P_m P_n 
+ \sqrt{\lambda} G_{mn} \mathcal{D}_1 X^m \mathcal{D}_1 X^n,
\ee
\be
\Phi_1 = \Pi_{\mathfrak{a}} \mathcal{D}_1 X^{\mathfrak{a}} + 
P_m \mathcal{D}_1 X^m.
\ee
Here $\Pi_+ := P_+ - P_{\hat{\alpha}} 
\Lambda^{\hat{\alpha}}{}_{\hat{\beta}} \Psi^{\hat{\beta}}$
and
$\Pi_-:= P_-$.

Since the action \eqref{KFGS} is a singular system,
it is necessary to 
introduce fermionic Lagrange multipliers $\chi^{\hat{\alpha}}$ 
for \eqref{PCF}.
The Hamilton form of the equations of motion are given by
\bel{dtZ}
\dot{Z}^M(\tau,\sigma) = \{ Z^M(\tau,\sigma), H \}_{\mathrm{P.B.}}
+ \frac{1}{2\pi} \int \de \sigma' \, \{ Z^M(\tau, \sigma),
\Phi_{\hat{\alpha}}(\tau, \sigma') \}_{\mathrm{P.B.}} \, 
\chi^{\hat{\alpha}}( \tau, \sigma'),
\ee
\be
\dot{P}_M(\tau,\sigma) = \{ P_M(\tau,\sigma), H \}_{\mathrm{P.B.}}
+ \frac{1}{2\pi} \int \de \sigma' \, \{ P_M(\tau, \sigma),
\Phi_{\hat{\alpha}}(\tau, \sigma') \}_{\mathrm{P.B.}} \, 
\chi^{\hat{\alpha}}( \tau, \sigma'),
\ee
where
\be
H = \frac{1}{2\pi} \int \de \sigma \, \mathcal{H}.
\ee

The singularity of the action comes from the fact that
$\dot{\Psi}^{\hat{\alpha}}$ or 
equivalently $\mathcal{D}_0 \Psi^{\hat{\alpha}}$
does not appear in \eqref{conjM}; 
$\dot{\Psi}^{\hat{\alpha}}$ can not be expressed by
the phase space variables. 
The equations of motion \eqref{dtZ} for $Z^M = \Psi^{\hat{\alpha}}$
can be rewritten as $\mathcal{D}_0 \Psi^{\hat{\alpha}}
= \chi^{\hat{\alpha}}$.
Therefore, the introduction of the fermionic Lagrange multipliers
 $\chi^{\hat{\alpha}}$
is eventually equivalent to converting $\mathcal{D}_0 \Psi^{\hat{\alpha}}$
into $\chi^{\hat{\alpha}}$.


\subsection{Generalized light-cone gauge}

We first reduce the phase space from 
$\Gamma = \{ (X^{\ul{m}}, P_{\ul{m}}, 
\Psi^{\hat{\alpha}}, P_{\hat{\alpha}})\}$
to $\Gamma^* = \{(X^m, P_m, \Psi^{\hat{\alpha}}, P_{\hat{\alpha}})\}$
by taking the generalized light-cone gauge and by
solving the Virasoro constraints $\Phi_0 = 0$ and $\Phi_1=0$.
 
Let us take the generalized light-cone gauge:
\be
X^+ = \kappa \tau, \qq
\dot{P}_- = 0.
\ee
The Virasoro constraint $\Phi_1= 0$
is solved by setting 
$\mathcal{D}_1 X^- = - (1/P_-) P_m \mathcal{D}_1 X^m$.
The bosonic Lagrange multipliers are determined as
\be
h^{00} = - \frac{1}{\sqrt{\lambda} \kappa}
( G^{+ +} \Pi_+ + G^{+-} P_- ),
\qq
h^{01} = \frac{\sqrt{\lambda}}{P_-^2} G_{--} P_m 
\mathcal{D}_1 X^m.
\ee
The Virasoro constraint $\Phi_0=0$ yields 
the following quadratic equation
\be
\begin{split}
& ( G^{++} \Pi_+^2 + 2 G^{+-} P_- \, \Pi_+ + G^{--}P_-^2)
+ G^{mn} P_m P_n \cr
& + \frac{\lambda}{P_-^2} G_{--} ( P_m \mathcal{D}_1 X^m )^2
+ \lambda G_{mn} \mathcal{D}_1 X^m \mathcal{D}_1 X^n = 0,
\end{split}
\ee
which gives a solution
\be
P_+ = P_{\hat{\alpha}} \Lambda^{\hat{\alpha}}{}_{\hat{\beta}}
\Psi^{\hat{\beta}} + \Pi_+^{(\mathrm{sol})}.
\ee
Here
\be
\Pi_+^{(\mathrm{sol})} = 
\frac{1}{G^{++}}\left(
 - B + \epsilon_B \sqrt{B^2 - G^{++} \tilde{C}}
\right),
\ee
with $B = G^{+-}P_-$, $\epsilon_B = \mathrm{sign}(B)$,
\be
\begin{split}
\tilde{C}
&= G^{--}P_-^2 + \frac{1}{G_z} \sum_{a=1}^4 P_a^2 + \frac{1}{G_y}
\sum_{s=1}^4 P_{4+s}^2 \cr
& + \frac{\lambda}{P_-^2} G_{--}
( P_a \mathcal{D}_1 z^a + P_{4+s} \mathcal{D}_1 y^s)^2
+ \lambda G_z \sum_{a=1}^4 (\mathcal{D}_1 z^a)^2
+ \lambda G_y \sum_{s=1}^4 ( \mathcal{D}_1 y^s)^2.
\end{split}
\ee

The time evolution for the reduced phase variables is given by
\be
\dot{F}(\tau, \sigma) = \{ F(\tau,\sigma), 
H_{\mathrm{LC}} \}_{\mathrm{P.B.}}^*
+ \frac{1}{2\pi} \int \de \sigma' 
\{ F(\tau, \sigma), \Phi_{\hat{\alpha}}(\tau, \sigma')
 \}_{\mathrm{P.B.}}^*
\, \chi^{\hat{\alpha}}(\tau, \sigma').
\ee
Here $\{F,G\}_{\mathrm{P.B.}}^*$ is the Poisson bracket in the reduced
phase space $\Gamma^*$.
The light-cone Hamiltonian is given by
\be
H_{\mathrm{LC}} = - \frac{\kappa}{2\pi}\int_{-\pi}^{\pi} \de \sigma\, P_+,
\ee
and $\Phi_{\hat{\alpha}}$
is the fermionic constraints in the reduced phase space:
\be
\Phi_{\hat{\alpha}}= P_{\hat{\alpha}}
+ \sqrt{\lambda} B_{\hat{\alpha} \hat{\beta}}
\partial_1 \Psi^{\hat{\beta}} - P_- \Lambda^-{}_{\hat{\alpha}}
- P_m \Lambda^m{}_{n \hat{\alpha}} X^n.
\ee
As in the bosonic case, the light-cone Hamiltonian can be understood
by canonical transformation. It can be also explained 
by using the first order form of the action:
\be
S = \frac{1}{2\pi} \int \de^2 \xi
\left(
P_+ \dot{X}^+ + P_- \dot{X}^-
+ P_{m} \dot{X}^{m}
+ P_{\hat{\alpha}} \dot{\Psi}^{\hat{\alpha}}
- \mathcal{H} - \Phi_{\hat{\alpha}} \chi^{\hat{\alpha}} \right).
\ee
By taking the generalized light-cone gauge and
by substituting the solutions of the Virasoro constraints
into the action,
we have
\be
S = \frac{1}{2\pi} \int \de^2 \xi
\left( P_m \dot{X}^m + P_{\hat{\alpha}} \dot{\Psi}^{\hat{\alpha}}
- \mathcal{H}_{\mathrm{LC}} - 
\Phi_{\hat{\alpha}} \chi^{\hat{\alpha}} \right).
\ee
Here $\mathcal{H}_{\mathrm{LC}}= - \kappa P_+$ and
we have dropped the total $\tau$-derivative term $P_- \dot{X}^-$.

We can see that the remaining fermionic constraints 
$\Phi_{\hat{\alpha}} \approx 0$ are second class:
\be
\{ \Phi_{\hat{\alpha}}(\tau, \sigma),
\Phi_{\hat{\beta}}(\tau, \sigma') \}_{\mathrm{P.B.}}^*
= -2 \pi \mathcal{C}_{\hat{\alpha}\hat{\beta}}(\tau,\sigma)
\delta(\sigma - \sigma'),
\ee
where
\be
\begin{split}
\mathcal{C}_{\hat{\alpha} \hat{\beta}}
&= P_- ( \partial \Lambda^-{}_{\hat{\alpha}}/ 
\partial \Psi^{\hat{\beta}})
+ P_- ( \partial \Lambda^-{}_{\hat{\beta}}/ 
\partial \Psi^{\hat{\alpha}}) \cr
& - P_m X^n \left(
\Lambda^m{}_{k \hat{\alpha}} \Lambda^k{}_{n \hat{\beta}}
+ \Lambda^m{}_{k \hat{\beta}} \Lambda^k{}_{n \hat{\alpha}}
- ( \partial \Lambda^m{}_{n\hat{\alpha}}/
\partial \Psi^{\hat{\beta}} )
- ( \partial \Lambda^m{}_{n\hat{\beta}}/
\partial \Psi^{\hat{\alpha}} ) \right) 
+ \sqrt{\lambda} (\partial_1 B_{\hat{\alpha} \hat{\beta}}) \cr
& - \sqrt{\lambda} \left[
( \partial_m B_{\hat{\alpha} \hat{\gamma}} )
\Lambda^m{}_{k \hat{\beta}} X^k
+ ( \partial_m B_{\hat{\beta} \hat{\gamma}} )
\Lambda^m{}_{k \hat{\alpha}} X^k 
- (\partial B_{\hat{\alpha} \hat{\gamma}}/
\partial \Psi^{\hat{\beta}} )
- (\partial B_{\hat{\beta} \hat{\gamma}}/
\partial \Psi^{\hat{\alpha}} )
\right] \partial_1 \Psi^{\hat{\gamma}}. \nonumber
\end{split}
\ee
We assume that $\mathcal{C}$ is invertible.
For $AdS_5 \times S^5$, this is indeed the case since
the terms in the first line of the above equation
start with an invertible matrix:
\be
P_- ( \partial \Lambda^-{}_{\hat{\alpha}}/ 
\partial \Psi^{\hat{\beta}})
+ P_- ( \partial \Lambda^-{}_{\hat{\beta}}/ 
\partial \Psi^{\hat{\alpha}})
= 2 \sqrt{2} \, \im \, 
P_- \, ( \gamma_+ )_{\hat{\alpha} \hat{\beta}} 
+ \mathcal{O}(\Psi^2),
\ee
\be
( \gamma_+ )_{\hat{\alpha} \hat{\beta}}
= \mat{ 0 & \ & ( \gamma_+ )_{\alpha \bar{\beta}}  \cr
( \gamma_+)_{\bar{\alpha} \beta} & & 0 }
= \mat{ 0 & \ & 1_8 \cr 1_8 & & 0 }.
\ee
The consistency of the time evolution of the fermionic
constraints $(\dot{\Phi}_{\hat{\alpha}}=0)$
determines the fermionic Lagrange multipliers as follows
\be
\chi^{\hat{\alpha}}(\tau, \sigma)
= \bigl( \mathcal{C}^{-1}(\tau,\sigma) 
\bigr)^{\hat{\alpha} \hat{\beta}} 
\mathcal{X}_{\hat{\beta}}(\tau, \sigma).
\ee
Here $\mathcal{X}_{\hat{\beta}}(\tau, \sigma)
= \{ \Phi_{\hat{\beta}}(\tau,\sigma), H_{\mathrm{LC}} 
\}_{\mathrm{P.B.}}^*$. 
Since the explicit form of $\mathcal{X}_{\hat{\beta}}$ 
is rather lengthy and is not necessary here, we do not
write it in this paper.

The (equal $\tau$) Dirac bracket is given by
\be
\{ F, G \}_{\mathrm{D.B.}}
= \{ F, G \}_{\mathrm{P.B.}}^*
+ \frac{1}{2\pi} \int \de \sigma
\{ F, \Phi_{\hat{\alpha}}(\tau,\sigma) \}_{\mathrm{P.B.}}^*\, 
\bigl(\mathcal{C}^{-1}(\tau,\sigma)\bigr)^{\hat{\alpha} \hat{\beta}}
\, \{ \Phi_{\hat{\beta}}(\tau, \sigma), G \}_{\mathrm{P.B.}}^*.
\ee
Using the Dirac bracket we can choose $(X^m, P_m, \Psi^{\hat{\alpha}})$
as dynamical variables and $P_{\hat{\alpha}}$ can be
treated as the solution of the fermionic constraints:
\be
P_{\hat{\alpha}} = - \sqrt{\lambda} 
B_{\hat{\alpha} \hat{\beta}} \partial_1 \Psi^{\hat{\beta}}
+ P_- \Lambda^-{}_{\alpha} + P_m \Lambda^m{}_{n\hat{\alpha}}
X^n.
\ee

The time evolution of the dynamical variables are now given by
\be
\dot{F} = \{ F, H_{\mathrm{LC}} \}_{\mathrm{D.B.}}.
\ee

Let
\be
\mathcal{U}^m{}_{\hat{\alpha}} = \Lambda^m{}_{n \hat{\alpha}} X^n,
\qq
\mathcal{V}_{m \hat{\alpha}}
= \sqrt{\lambda} ( \partial_m B_{\hat{\alpha} \hat{\beta}} )
\partial_1 \Psi^{\hat{\beta}} - P_n \Lambda^n{}_{m \hat{\alpha}}.
\ee
The Dirac bracket among the dynamical variables are given by 
\be
\begin{split}
\{ X^m(\tau, \sigma), X^n(\tau, \sigma') \}_{\mathrm{D.B.}}
&= -2\pi \, \mathcal{U}^m{}_{\hat{\alpha}}
\bigl( \mathcal{C}^{-1} \bigr)^{\hat{\alpha} \hat{\beta}}
\mathcal{U}^n{}_{\hat{\beta}}\, 
\delta( \sigma - \sigma'), \cr
\{ X^m(\tau, \sigma), P_n(\tau, \sigma') \}_{\mathrm{D.B.}} 
&= 2\pi \left( \delta^m_n 
- \mathcal{U}^m{}_{\hat{\alpha}} 
\bigl( \mathcal{C}^{-1}\bigr)^{\hat{\alpha} \hat{\beta}}
\mathcal{V}_{n \hat{\beta}} \right) \, \delta(\sigma - \sigma'),  \cr
\{ X^m(\tau, \sigma), \Psi^{\hat{\alpha}}(\tau, \sigma') \}_{\mathrm{D.B.}}
&= - 2\pi\, \mathcal{U}^m{}_{\hat{\beta}} \bigl(
\mathcal{C}^{-1} \bigr)^{\hat{\beta} \hat{\alpha}}
\delta(\sigma - \sigma'), \cr
\{ P_m(\tau, \sigma), P_n(\tau, \sigma')\}_{\mathrm{D.B.}}
&= - 2\pi \, \mathcal{V}_{m \hat{\alpha}}
\bigl( \mathcal{C}^{-1} \bigr)^{\hat{\alpha} \hat{\beta}}
\mathcal{V}_{n \hat{\beta}}\, \delta(\sigma - \sigma'), \cr
\{ P_m(\tau, \sigma), \Psi^{\hat{\alpha}}(\tau, \sigma') \}_{\mathrm{D.B.}}
&= - 2\pi\, \mathcal{V}_{m \hat{\beta}}
\bigl( \mathcal{C}^{-1} \bigr)^{\hat{\beta} \hat{\alpha}} \,
\delta( \sigma - \sigma'), \cr
\{ \Psi^{\hat{\alpha}}(\tau, \sigma),
\Psi^{\hat{\beta}}(\tau, \sigma') \}_{\mathrm{D.B.}}
&= 2 \pi \bigl( \mathcal{C}^{-1}
\bigr)^{\hat{\alpha} \hat{\beta}}
\delta( \sigma - \sigma').
\end{split}
\ee
The quantization of these transverse degrees of
freedom is then a straightforward task: 
to replace $\im \hbar$ times the Dirac bracket by
the graded commutator.
Because of the fermionic constraints,
all corresponding quantum operators become
non-commutative.

\vspace{0.3cm}


\noindent
{\bf{Acknowledgements}}

We would like to thank Makoto Sakaguchi and Kentaroh Yoshida 
for useful discussions. 
This work is supported by the 21 COE program
``Construction of wide-angle mathematical basis focused on knots"
and in part by the Grant-in Aid for Scientific
Research (No. 18540285)
from Japan Ministry of Education.


\appendix

\section{Details on the induced vielbein}

The $psu(2,2|4)$ generators are given by
\be
\widehat{P}_{\ul{a}} 
= ( \widehat{P}_{\hat{a}}, \widehat{P}_{\hat{a}'}), \qq
\widehat{J}_{\hat{a}\hat{b}} 
= - \widehat{J}_{\hat{b} \hat{a}}, \qq
\widehat{J}_{\hat{a}' \hat{b}'}
= - \widehat{J}_{\hat{b}' \hat{a}' },
\qq
\widehat{Q}_{\ul{\alpha}}, \qq
\widehat{\ol{Q}}_{\ul{\bar{\alpha}}},
\ee
where $\ul{a}=0,1,2,\dotsc, 9$,
$\hat{a}, \hat{b}=0,1,2,3,4$, $\hat{a}', \hat{b}' = 5,6,7,8,9$,
$\ul{\alpha}=1,2,\dotsc, 16$, $\ul{\bar{\alpha}}=\bar{1}, \bar{2},
\dotsc, \bar{16}$.
The bosonic generatros are chosen to be anti-Hermitian
and $(\widehat{Q}_{\ul{\alpha}})^{\dag} 
= \widehat{\ol{Q}}_{\ul{\bar{\alpha}}}$.
The non-zero commutation relations are given by
\begin{align}
[\widehat{P}_{\hat{a}}, \widehat{P}_{\hat{b}} ] 
&= \widehat{J}_{\hat{a}\hat{b}},&
[\widehat{P}_{\hat{a}'}, \widehat{P}_{\hat{b}'} ] 
&= - \widehat{J}_{\hat{a}'\hat{b}'}, \\
[\widehat{P}_{\hat{a}}, \widehat{J}_{\hat{b}\hat{c}} ] 
&= \eta_{\hat{a}\hat{b}} \widehat{P}_{\hat{c}} 
- \eta_{\hat{a}\hat{c}} \widehat{P}_{\hat{b}},&
[\widehat{P}_{\hat{a}'}, \widehat{J}_{\hat{b}'\hat{c}'} ] 
&= \delta_{\hat{a}'\hat{b}'} 
\widehat{P}_{\hat{c}'} 
- \delta_{\hat{a}'\hat{c}'} \widehat{P}_{\hat{b}'}, \\
[ \widehat{J}_{\hat{a}\hat{b}}, \widehat{J}_{\hat{c}\hat{d}} ] 
&= \eta_{\hat{b}\hat{c}} 
\widehat{J}_{\hat{a}\hat{d}} + \mbox{$3$ terms},&
[ \widehat{J}_{\hat{a}'\hat{b}'}, \widehat{J}_{\hat{c}'\hat{d}'} ] 
&= \delta_{\hat{b}'\hat{c}'} \widehat{J}_{\hat{a}'\hat{d}'}
+ \mbox{$3$ terms}, \\
[ \widehat{Q}_{\ul{\alpha}}, \widehat{P}_{\ul{a}} ]
&= \frac{\im}{2} 
( \gamma_{\ul{a}} \varrho )_{\ul{\alpha}}{}^{\ul{\beta}} 
\widehat{Q}_{\ul{\beta}},&
[ \widehat{\ol{Q}}_{\ul{\bar{\alpha}}}, \widehat{P}_{\ul{a}} ]
&= - \frac{\im}{2} 
( \gamma_{\ul{a}} \varrho )_{\ul{\bar{\alpha}}}
{}^{\ul{\bar{\beta}}} 
\widehat{\ol{Q}}_{\ul{\bar{\beta}}},\\
[ \widehat{Q}_{\ul{\alpha}}, \widehat{J}_{\hat{a}\hat{b}}] 
&= \frac{1}{2} 
( \gamma_{\hat{a}\hat{b}} )_{\ul{\alpha}}{}^{\ul{\beta}} 
\widehat{Q}_{\ul{\beta}}, &
[ \widehat{Q}_{\ul{\alpha}}, \widehat{J}_{\hat{a}'\hat{b}'} ] 
&= \frac{1}{2} 
( \gamma_{\hat{a}'\hat{b}'} )_{\ul{\alpha}}{}^{\ul{\beta}}
\widehat{Q}_{\ul{\beta}}, \\
[ \widehat{\ol{Q}}_{\ul{\bar{\alpha}}}, 
\widehat{J}_{\hat{a}\hat{b}} ] &=
\frac{1}{2} ( \gamma_{\hat{a}\hat{b}})_{\ul{\bar{\alpha}}}
{}^{\ul{\bar{\beta}}} \widehat{\ol{Q}}_{\ul{\bar{\beta}}},&
[ \widehat{\ol{Q}}_{\ul{\bar{\alpha}}}, 
\widehat{J}_{\hat{a}'\hat{b}'} ]
&= \frac{1}{2} 
( \gamma_{\hat{a}'\hat{b}'})_{\ul{\bar{\alpha}}}
{}^{\ul{\bar{\beta}}} \widehat{\ol{Q}}_{\ul{\bar{\beta}}},
\end{align}
\be
\{ \widehat{Q}_{\ul{\alpha}}, 
\widehat{\ol{Q}}_{\ul{\bar{\beta}}} \}
= - 2 \im ( \gamma^{\ul{a}} )_{\ul{\alpha} \ul{\bar{\beta}}} 
\widehat{P}_{\ul{a}}
+ ( \gamma^{\hat{a}\hat{b}} \varrho)_{\ul{\alpha} \ul{\bar{\beta}}} 
\widehat{J}_{\hat{a}\hat{b}}
- ( \gamma^{\hat{a}'\hat{b}'} 
\varrho)_{\ul{\alpha} \ul{\bar{\beta}}}
 \widehat{J}_{\hat{a}'\hat{b}'}.
\ee
Here $\eta_{\hat{a}\hat{b}} =\mathrm{diag}(-,+,+,+,+)$.
We define
\be
\widehat{P}_{\pm} = \frac{1}{\sqrt{2}}( \widehat{P}_0
\pm \widehat{P}_9 ), \qq
\gamma_{\pm} = \frac{1}{2} ( \gamma_0 \pm \gamma_ 9).
\ee
\be
( \gamma_+ )_{\ul{\alpha} \ul{\beta}}
= \mat{ ( \gamma_+ )_{\alpha \beta} & \ & 0 \cr
0 & & 0 } = \mat{ 1_8 & \ & 0 \cr 0 & & 0}, \qq
( \gamma_-)_{\ul{\alpha} \ul{\beta}}
= \mat{ 0 & \ & 0 \cr 0 & & ( \gamma_-)_{\dot{\alpha} \dot{\beta}} }
= \mat{ 0 & \ & 0 \cr 0 & & 1_8 }.
\ee
In our notation, 
\be
(\gamma^a)_{\ul{\alpha} \ul{\beta}}
= \mat{ 0 & \ & ( \gamma^a)_{\alpha \dot{\beta}} \cr
( \gamma^a)_{\dot{\alpha} \beta} & & 0 }, \qq
(\gamma^{4+s})_{\ul{\alpha} \ul{\beta}}
= \mat{ 0 & \ & ( \gamma^{4+s})_{\alpha \dot{\beta}} \cr
( \gamma^{4+s})_{\dot{\alpha} \beta} & & 0 }, 
\ee
for $a=1,2,3,4$ and $s=1,2,3,4$.

If we decompose the fermionic generators
into $\widehat{Q}_{\ul{\alpha}} = ( \widehat{Q}_{\alpha}^+,
\widehat{Q}_{\dot{\alpha}}^-)$,
$\widehat{\ol{Q}}_{\ul{\bar{\alpha}}}
= ( \widehat{\ol{Q}}{}_{\bar{\alpha}}^+,
\widehat{\ol{Q}}{}_{\dot{\bar{\alpha}}}^-)$,
some of commutation relations can be rewritten as follows:
\begin{align}
[ \widehat{Q}_{\alpha}^+, 
\widehat{P}_+ ] &= \frac{\im}{\sqrt{2}} 
( \gamma_+ \varrho )_{\alpha}{}^{\beta}
\widehat{Q}_{\beta}^+, &
[ \widehat{Q}_{\dot{\alpha}}^-, \widehat{P}_+ ] &= 0, \\
[ \widehat{Q}_{\alpha}^+, \widehat{P}_- ] &= 0, &
[ \widehat{Q}_{\dot{\alpha}}^-, \widehat{P}_- ] 
&= \frac{\im}{\sqrt{2}} 
( \gamma_- \varrho)_{\dot{\alpha}}{}^{\dot{\beta}}
\widehat{Q}_{\dot{\beta}}^-,
\end{align}
\be
\{ \widehat{Q}_{\alpha}^+, \widehat{\ol{Q}}{}_{\bar{\beta}}^+ \}
= 2 \sqrt{2} \im \, (\gamma_+)_{\alpha \bar{\beta}} \, 
\widehat{P}_-
+ ( \gamma^{ab} \varrho)_{\alpha \bar{\beta}} \,
\widehat{J}_{ab} - ( \gamma^{a'b'} \varrho)_{\alpha \bar{\beta}}
\, \widehat{J}_{a'b'}.
\ee
Here $a,b=1,2,3,4$, $a',b'=5,6,7,8$.

Using the coset representative \eqref{CRG} with
\begin{align}
g_z &= \exp\left( \mathcal{X}^a \widehat{P}_a \right),&
\mathcal{X}^a &= \frac{z^a}{z} 
\log \left( \frac{1+(1/2)z }{1 - (1/2)z} \right), \\
g_y &= \exp\left( \mathcal{X}^{4+s} \widehat{P}_{4+s} \right),&
\mathcal{X}^{4+s} &= - \im \frac{y^s}{y} \log
\left( \frac{1 + (\im/2) y  }{1 - (\im/2) y} \right),
\end{align}
we can calculate the vielbeins for the reduced type IIB
superspace as follows:
\be
G^{-1} \de G
= E^{\ul{a}} \widehat{P}_{\ul{a}}
+ E^{\ul{\alpha}} \widehat{Q}_{\ul{\alpha}}
+ \ol{E}^{\ul{\bar{\alpha}}} \widehat{\ol{Q}}_{\ul{\bar{\alpha}}}
+ (\mbox{spin connection part}).
\ee

Let us define a $16 \times 16$ matrix $\mathcal{M}^2$
by
\be
\mathcal{M}^2 = \mat{
(\mathcal{M}^2)^{\alpha}{}_{\beta} & \ &
(\mathcal{M}^2)^{\alpha}{}_{\bar{\beta}} \cr
(\mathcal{M}^2)^{\bar{\alpha}}{}_{\beta} & \ &
(\mathcal{M}^2)^{\bar{\alpha}}{}_{\bar{\beta}} },
\ee
where the matrix elements are defined by
\be
\mathrm{ad}^2
( \theta^+ \widehat{Q}^+ 
+ \bar{\theta}^+ \widehat{\ol{Q}}{}^+ )
( \widehat{Q}_{\alpha}^+) 
= \widehat{Q}_{\beta}^+ ( \mathcal{M}^2)^{\beta}{}_{\alpha} 
+ \widehat{\ol{Q}}
{}_{\bar{\beta}}^+ ( \mathcal{M}^2 )^{\bar{\beta}}{}_{\alpha}, 
\ee
\be
\mathrm{ad}^2
( \theta^+ \widehat{Q}^+ 
+ \bar{\theta}^+ \widehat{\ol{Q}}{}^+ )
( \widehat{\ol{Q}}_{\bar{\alpha}}\!\!\!{}^+) 
= 
\widehat{Q}_{\beta}^+ ( \mathcal{M}^2)^{\beta}{}_{\bar{\alpha}} 
+ \widehat{\ol{Q}}{}_{\bar{\beta}}^+ 
( \mathcal{M}^2 )^{\bar{\beta}}{}_{\bar{\alpha}}.
\ee
Explicit form of the matrix elements are given by
\be
\begin{split}
( \mathcal{M}^2)^{\alpha}{}_{\beta}
&= \frac{1}{2} ( \theta^+ \gamma_{ab} )^{\alpha}
( \bar{\theta}^+ \gamma^{ab} \varrho)_{\beta}
- \frac{1}{2} ( \theta^+ \gamma_{a'b'} )^{\alpha}
( \bar{\theta}^+ \gamma^{a'b'} \varrho)_{\beta}, \cr
( \mathcal{M}^2)^{\alpha}{}_{\bar{\beta}}
&= - \frac{1}{2} ( \theta^+ \gamma_{ab} )^{\alpha}
( \theta^+ \gamma^{ab} \varrho)_{\bar{\beta}}
+ \frac{1}{2} ( \theta^+ \gamma_{a'b'} )^{\alpha}
( \theta^+ \gamma^{a'b'} \varrho)_{\bar{\beta}}, \cr
( \mathcal{M}^2)^{\bar{\alpha}}{}_{\beta}
&= \frac{1}{2} ( \bar{\theta}^+ \gamma_{ab} )^{\bar{\alpha}}
( \bar{\theta}^+ \gamma^{ab} \varrho)_{\beta}
- \frac{1}{2} ( \bar{\theta}^+ \gamma_{a'b'} )^{\bar{\alpha}}
( \bar{\theta}^+ \gamma^{a'b'} \varrho)_{\beta}, \cr
( \mathcal{M}^2)^{\bar{\alpha}}{}_{\bar{\beta}}
&= - \frac{1}{2} ( \bar{\theta}^+ \gamma_{ab} )^{\bar{\alpha}}
( \theta^+ \gamma^{ab} \varrho)_{\bar{\beta}}
+ \frac{1}{2} ( \bar{\theta}^+ \gamma_{a'b'} )^{\bar{\alpha}}
( \theta^+ \gamma^{a'b'} \varrho)_{\bar{\beta}}.
\end{split}
\ee
Let
\be
\frac{\cosh \mathcal{M}-1_{16}}{\mathcal{M}^2}
= \mat{ (K_{11})^{\alpha}{}_{\beta} & \ & 
(K_{12})^{\alpha}{}_{\bar{\beta}} \cr
(K_{21})^{\bar{\alpha}}{}_{\beta} & &
(K_{22})^{\bar{\alpha}}{}_{\bar{\beta}} },
\qq
\frac{\sinh \mathcal{M}}{\mathcal{M}}
= \mat{ (L_{11})^{\alpha}{}_{\beta} & \ & 
(L_{12})^{\alpha}{}_{\bar{\beta}} \cr
(L_{21})^{\bar{\alpha}}{}_{\beta} & &
(L_{22})^{\bar{\alpha}}{}_{\bar{\beta}} }.
\ee
The induced vielbeins are calculated as follows:
\be
\begin{split}
E^{\pm}_i &= e^{\pm}{}_+ \partial_i X^+ + e^{\pm}{}_- 
\mathcal{D}_i X^- , \cr
E^a_i &= \frac{1}{1 - (z^2/4) } \mathcal{D}_i z^a, \cr
E^{4+s}_i &= \frac{1}{1 + (y^2/4)}
\mathcal{D}_i y^s, \cr
E^{+\alpha}_i
&=  U^{\alpha}{}_{\beta}
\left( (L_{11} \mathcal{D}_i \theta^+)^{\beta} 
+ (L_{12} \mathcal{D}_i \bar{\theta}^+ )^{\beta} \right), \cr
E^{- \dot{\alpha}}_i
&= V^{\dot{\alpha}}{}_{\beta}
\left( (L_{11} \mathcal{D}_i \theta^+)^{\beta} 
+ (L_{12} \mathcal{D}_i \bar{\theta}^+ )^{\beta} \right), \cr
\ol{E}^{+\bar{\alpha}}_i
&= \ol{U}^{\bar{\alpha}}{}_{\bar{\beta}}
\left( 
(L_{21} \mathcal{D}_i \theta^+)^{\bar{\beta}}
+ ( L_{22} \mathcal{D}_i \bar{\theta}^+)^{\bar{\beta}} 
\right), \cr
\ol{E}^{- \dot{\bar{\alpha}}}_i
&= \ol{V}^{\dot{\bar{\alpha}}}{}_{\bar{\beta}}
\left( 
(L_{21} \mathcal{D}_i \theta^+)^{\bar{\beta}}
+ ( L_{22} \mathcal{D}_i \bar{\theta}^+)^{\bar{\beta}} 
\right), 
\end{split}
\ee
where
\be
\begin{split}
e^{\pm}{}_+
&= \frac{1}{2} 
\left[ 
\left( \frac{1 + (z^2/4)}{1 - (z^2/4)} \right) \pm 
\left( \frac{1 - (y^2/4)}{1 + (y^2/4)} \right)
\right], \cr
e^{\pm}{}_-
&=\frac{1}{2} 
\left[ 
\left( \frac{1 + (z^2/4)}{1 - (z^2/4)} \right) \mp 
\left( \frac{1 - (y^2/4)}{1 + (y^2/4)} \right)
\right],
\end{split}
\ee
\be
U^{\alpha}{}_{\beta}
= \frac{
\left( \delta^{\alpha}{}_{\beta}
+ (1/4) z^a y^s ( \gamma_a \gamma_{4+s} )^{\alpha}{}_{\beta} 
\right)}
{\left( 1 - (z^2/4) \right)^{1/2}
\left( 1 + (y^2/4) \right)^{1/2} },
\qq
\ol{U}^{\bar{\alpha}}{}_{\bar{\beta}}
= \frac{
\left( \delta^{\bar{\alpha}}{}_{\bar{\beta}}
+ (1/4) z^a y^s 
( \gamma_a \gamma_{4+s} )^{\bar{\alpha}}{}_{\bar{\beta}} 
\right)}
{\left( 1 - (z^2/4) \right)^{1/2}
\left( 1 +  (y^2/4) \right)^{1/2} },
\ee
\be
V^{\dot{\alpha}}{}_{\beta}
= \frac{
- \im z^a ( \gamma_a \varrho)^{\dot{\alpha}}{}_{\beta}
+ \im y^s ( \gamma_{4+s} \varrho)^{\dot{\alpha}}{}_{\beta}}
{ 2 \left( 1 - (z^2/4) \right)^{1/2}
\left(1 + (y^2/4) \right)^{1/2} },
\qq
\ol{V}^{\dot{\bar{\alpha}}}{}_{\bar{\beta}}
= \frac{
 \im z^a 
( \gamma_a \varrho)^{\dot{\bar{\alpha}}}{}_{\bar{\beta}}
- \im y^s 
( \gamma_{4+s} \varrho)^{\dot{\bar{\alpha}}}{}_{\bar{\beta}}}
{2 \left( 1 - (z^2/4) \right)^{1/2}
\left(1 + (y^2/4) \right)^{1/2} },
\ee
\be
\begin{split}
\mathcal{D}_i X^-
&= \partial_i X^- 
+ 2 \sqrt{2} \im \left[
( \bar{\theta}^+ \gamma_+ K_{11} )_{\alpha}
+ ( \theta^+ \gamma_+ K_{21} )_{\alpha} \right]
\mathcal{D}_i \theta^{+\alpha} \cr
& \qq \qq
+ 2 \sqrt{2} \im \left[
( \bar{\theta}^+ \gamma_+ K_{12} )_{\bar{\alpha}}
+ ( \theta^+ \gamma_+ K_{22} )_{\bar{\alpha}}
\right] \mathcal{D}_i \bar{\theta}^{+\bar{\alpha}}, \cr
\mathcal{D}_i z^a &= \partial_i z^a
- 2 z_b \left[
( \bar{\theta}^+ \gamma^{ab} \varrho K_{11} )_{\alpha}
- ( \theta^+ \gamma^{ab} \varrho K_{21} )_{\alpha}
\right] 
\mathcal{D}_i \theta^{+\alpha} \cr
& \qq \ \ \  - 2 z_b
\left[ ( \bar{\theta}^+ \gamma^{ab} \varrho
K_{12} )_{\bar{\alpha}}
- ( \theta^+ \gamma^{ab} \varrho K_{22} )_{\bar{\alpha}}
\right] \mathcal{D}_i \bar{\theta}^{+\bar{\alpha}}, \cr
\mathcal{D}_i y^s
&= \partial_i y^s
+ 2 y_{s'}
\left[
(\bar{\theta}^+ \gamma^{4+s,4+s'} \varrho K_{11} )_{\alpha}
- ( \theta^+ \gamma^{4+s,4+s'} \varrho K_{21} )_{\alpha}
\right]
\mathcal{D}_i \theta^{+\alpha} \cr
& \qq \ \ \ 
+ 2 y_{s'}
\left[
( \bar{\theta}^+ \gamma^{4+s,4+s'} \varrho K_{12} )_{\bar{\alpha}}
+ ( \theta^+ \gamma^{4+s,4+s'} \varrho K_{22} )_{\bar{\alpha}}
\right]
\mathcal{D}_i \bar{\theta}^{+\bar{\alpha}},  \cr
\mathcal{D}_i \theta^{+\alpha}
&=\partial_i \theta^{+\alpha} - \frac{\im}{\sqrt{2}}
( \theta^+ \gamma_+\, \varrho )^{\alpha} \partial_i X^+, \cr
\mathcal{D}_i \bar{\theta}^{+\bar{\alpha}}
&= \partial_i \bar{\theta}^{+\bar{\alpha}} + \frac{\im}{\sqrt{2}}
( \bar{\theta}^+ \gamma_+ \, \varrho)^{\bar{\alpha}} \partial_i X^+.
\end{split}
\ee
By comparing with \eqref{covD}, we can read off 
$\Lambda^-{}_{\hat{\alpha}}$, $\Lambda^m{}_{n \hat{\alpha}}$,
$\Lambda^{\hat{\alpha}}{}_{\hat{\beta}}$.
For example,
\be
\Lambda^-{}_{\alpha}
= 2 \sqrt{2} \, \im 
\left[
(\bar{\theta}^+ \gamma_+ K_{11} )_{\alpha}
+ ( \theta^+ \gamma_+ K_{21} )_{\alpha} \right],
\ee
\be
\Lambda^-{}_{\bar{\alpha}}
= 2 \sqrt{2} \, \im 
\left[
(\bar{\theta}^+ \gamma_+ K_{12} )_{\bar{\alpha}}
+ ( \theta^+ \gamma_+ K_{22} )_{\bar{\alpha}} \right].
\ee
The fields $B_{\hat{\alpha} \hat{\beta}}$
in the Wess-Zumino term are read off from
\be
\begin{split}
& \frac{1}{2} \epsilon^{ij} 
B_{\hat{\alpha} \hat{\beta}}(Z)
\mathcal{D}_i \Psi^{\hat{\alpha}} \mathcal{D}_j \Psi^{\hat{\beta}} \cr
&=
\epsilon^{ij} ( E^{+\alpha}_i \varrho_{\alpha \beta} E^{+\beta}_j
+ E^{-\dot{\alpha}}_i \varrho_{\dot{\alpha} \dot{\beta}}
E^{-\dot{\beta}}_j
- \ol{E}^{+\bar{\alpha}}_i \varrho_{\bar{\alpha} \bar{\beta}}
\ol{E}^{+\bar{\beta}}_j
- \ol{E}^{-\dot{\bar{\alpha}}}_i
\varrho_{\dot{\bar{\alpha}} \dot{\bar{\beta}}}
\ol{E}^{- \dot{\bar{\beta}}}_j ).
\end{split}
\ee
Our convention for the Levi-Civita symbol is $\epsilon^{01}=1$.


\end{document}